\newif\ifmnras
\mnrastrue
\ifmnras
	\documentclass[useAMS,usenatbib]{mn2e}
\else
	\documentclass[apj,numberedappendix]{emulateapj}
\fi
\usepackage{times}
\usepackage{graphics,epsf}
\usepackage{amsmath}                % American Mathematical Society package
\usepackage{amsfonts}               % American Mathematical Society fonts
\usepackage{amssymb}                % American Mathematical Society symbol
\usepackage{epsfig}                 % EPS figures
\usepackage{graphicx}
\usepackage{url}
\usepackage{tabularx}
\usepackage{booktabs}
\usepackage{subfigure}
\usepackage{graphicx}
\usepackage{grffile}
\def \cm{~\mathrm{cm}}
\def \s{~\mathrm{s}}

\def \km{~\mathrm{km}}
\def \kms{~\mathrm{km}~\mathrm{s}^{-1}}

\def \K{~\mathrm{K}}
\def \g{~\mathrm{g}}

\def \erg{~\mathrm{erg}}

\def \yr{~\mathrm{yr}}
\def \myr{~\mathrm{Myr}}

\def \kpc{~\mathrm{kpc}}

\def \tracerfour{\mathrm{tracer~A}}
\def \tracerthree{\mathrm{tracer~B}}
\usepackage{color}

\ifmnras
	\def \aap{A\&A}
	\def \araa{ARA\&A}
	\def \apj{ApJ}
	\def \apjl{ApJ}
	\def \apjs{ApJS}
	\def \nat{Nature}
	
	\def \mnras{MNRAS}

\fi

\ifmnras
	\title[Hitomi observations support heating by mixing]{Hitomi observations of Perseus support heating by mixing}
	\author[S. Hillel \& N. Soker]{Shlomi Hillel and Noam Soker \\
	Department of Physics, Technion -- Israel, Institute of Technology, Haifa 32000, Israel;
	shlomihi@tx.technion.ac.il;
	soker@physics.technion.ac.il}
\fi

\begin{document}

\ifmnras
	\pagerange{\pageref{firstpage}--\pageref{lastpage}} \pubyear{2016}

	\maketitle
\else
	\title{Hitomi observations of Perseus support heating by mixing}

	\author{Shlomi Hillel}
	\author{Noam Soker}
	\affil{Department of Physics, Technion -- Israel
	Institute of Technology, Haifa 32000, Israel;
	shlomihi@tx.technion.ac.il; soker@physics.technion.ac.il}
\fi

\label{firstpage}

\begin{abstract}
We compare the velocity dispersion of the intracluster medium (ICM) of the Perseus cluster of galaxies as observed by the Hitomi X-ray telescope to our three-dimensional hydrodynamical simulations of jet-inflated bubbles in cluster cooling flows, and conclude that the observations support the mixing-heating mechanism of the ICM. In the mixing-heating mechanism the ICM is heated by mixing of hot bubble gas with the ICM. This mixing is caused by vortices that are formed during the inflation process of the bubble. Sound waves and turbulence are also excited by the vortices, but they contribute less than 20 per cents to the heating of the ICM. Shocks that are excited by the jets contribute even less.
\smallskip
\textit{Key words:} galaxies: clusters: intracluster medium --- galaxies: clusters: individual: Perseus --- galaxies: jets
\end{abstract}

% ==========================================================
\section{INTRODUCTION}
\label{sec:introduction}
% ==========================================================

In its very short life time the Hitomi X-ray observatory has revealed extremely valuable information on the nature of the intracluster medium (ICM) of the Perseus cluster of galaxies \citep{Hitomi2016}. The analysis of Hitomi X-ray observations by \cite{Hitomi2016} show a  relatively quiescent ICM. The hot gas has a line-of-sight velocity dispersion of $164 \pm 10 \km \s^{-1}$ in the region $r=30 \kpc$ to $r= 60 \kpc$ from the central nucleus. This implies that the heating of the ICM by dissipation of turbulence energy cannot offset radiative cooling. Another heating mechanism is at work in the Perseus cluster.

The heating process of the ICM in cooling flows in clusters and groups of galaxies and in galaxies
operates in a negative feedback cycle (e.g., \citealt{Fabian2012, McNamaraNulsen2012, Farage2012, Gasparietal2013, Pfrommer2013, Soker2016}). The energy from the mass-accreting super massive black hole is carried to the ICM by jets (or collimated wind), hence it is termed the jet feedback mechanism (JFM). Several processes have been proposed to transfer the energy from the jets to the ICM (e.g., \citealt{Gasparietal2013}), some of which can act together, e.g., cosmic rays and thermal conduction (e.g., \citealt{GuoOh2008}).
\cite{Fabian2012} argues that sound waves, as inferred from the ripples in Perseus \citep{Fabianetal2006}, that are excited by jet-inflated bubbles \citep{SternbergSoker2009} are more efficient in transferring energy than weak shocks are.
\cite{Fabianetal2016} apply the sound wave heating mechanism to the Perseus cluster following the Hitomi observations.

\cite{Formanetal2007} propose that shocks heat the ICM in the Virgo cluster, while \cite{Randalletal2015} argue that shocks excited by periodic jets activity heat the ICM of the galaxy group NGC~5813. \cite{Sokeretal2016}, on the other hand, found problems in the process of shocks-heating of the ICM of NGC~5813. More generally, \cite{GilkisSoker2012} and \cite{HillelSoker2014} found in their two-dimensional numerical simulations, that mixing of the hot bubble gas with the ICM is much more significant than heating by shocks.
\cite{BruggenKaiser2002} and \cite{Bruggenetal2009} already discussed heating by mixing of the hot bubble gas with the ICM. However, their simulations were unrealistic as they injected artificial bubbles, and mixing was achieved by the destruction of the bubbles by instabilities.
The new simulations \citep{GilkisSoker2012, HillelSoker2014, HillelSoker2016} show the bubbles to survive for a long time, and show that jets induce vortices inside and outside the hot bubbles. These are the vortices that cause the efficient mixing of the hot bubble gas with the ICM.

Another heating mechanism is by dissipation of turbulence in the ICM (e.g., \citealt{DeYoung2010, Gasparietal2014}), or heating by turbulence and turbulent-mixing (e.g. \citealt{BanerjeeSharma2014}).
Based on X-ray observation \cite{Zhuravlevaetal2014} argue that dissipation of ICM turbulence is the main heating process of the ICM in the {Virgo} and {Perseus} cooling flow clusters.  \cite{Falcetaetal2010} deduce, based on numerical simulations, that turbulence cannot be the main heating source. \cite{Reynoldsetal2015}, based on idealized simulations with artificial bubble injection (namely, not with jet), reach a similar conclusion.

Cosmic rays can also heat the ICM (e.g., \citealt{Fujitaetal2013}), including the ICM in Perseus  (e.g., \citealt{FujitaOhira2013}).
\cite{Pfrommer2013} argues that mixing of cosmic rays that fill the jet-inflated bubbles with the ICM is essential for heating by cosmic rays.

In an earlier paper  \citep{HillelSoker2016} we conducted three-dimensional hydrodynamical simulations of jet-inflated bubbles in the ICM. For the parameters used in those simulations, we found the mixing by vortices to be the main heating mechanism (mixing-heating). We found that the mixing process accounts for $\simeq 80\%$ of the energy transferred from the jets to the ICM. Kinetic energy of the ICM carries about $20 \%$ of the energy that is deposited by the jets to the ICM. Only a fraction of the kinetic energy develops to ICM turbulence and sound waves. The other fraction of the kinetic energy is in large scale flows. Shocks play only a minor role in heating the ICM. Based on our earlier findings, as inferred from the fraction of kinetic energy, it seems that sound waves also play a minor role.
\cite{YangReynolds2016b} performed three-dimensional hydrodynamical simulations for a much longer time, and also found that mixing is the main heating process. They found though, that heating by shocks is the second important process and turbulent heating contributes a small fraction.
We note that turbulence can play a role in determining the properties of the ICM (e.g., \citealt{Gaspari2015}) even when it is not the main heating source.

In any case, turbulence in the ICM is expected as a by product of the mixing process that is induced by vortices \citep{HillelSoker2016, YangReynolds2016a}. This expectation from numerical simulations is compatible with findings of turbulence in the ICM of some cooling flows (e.g., \citealt{Zhuravlevaetal2014, Zhuravlevaetal2015, Arevalo2016, AndersonSunyaev2016, Hofmannetal2016}).

Following the results of Hitomi \citep{Hitomi2016} we conduct further analysis of our 3D numerical simulations. Although the simulations from \cite{HillelSoker2016} were not aiming at the Perseus cluster, the reanalysis of our simulations further strengthens our conclusion that heating by mixing process is the main heating process.

% ==========================================================
\section{SUMMARY OF NUMERICAL SETUP}
\label{sec:numerics}
% ==========================================================

We analyze the 3D hydrodynamical simulations that were presented in our earlier paper \citep{HillelSoker2016}, where more details on the numerical scheme and convergence tests can be found. We present here only the main features of the numerical code.

We use the {\sc pluto} code \citep{Mignone2007} for the
hydrodynamic simulations, in a three-dimensional Cartesian grid
with adaptive mesh refinement (AMR). The computational grid is in
the octant where the three coordinates $x$, $y$ and $z$ are
positive. At the $x = 0$, $y = 0$ and $z = 0$ planes we apply
reflective boundary conditions. The $z$ coordinate is chosen along
the initial axis of the jets. In reality two opposite jets are
launched simultaneously, such that the flow here is assumed to be
symmetric with respect to the $z = 0$ plane, amounting to
reflective boundary conditions at $z = 0$. The base computational
grid (lowest AMR level) spans the cube $0 \leq x, y, z \leq 50 \kpc$, with $16$
divisions in each direction.
Up to $5$ AMR levels are employed with a refinement ratio of $2$. Thus, the highest resolution is $\approx 0.1 \kpc$.

On the outer boundaries we use an outflow boundary condition.
At the boundary $z = 0$ we inject into the grid a jet with a
half-opening angle of $\theta_{\rm j} = 70^\circ$. The jet material is inserted through a
circle $x^2 + y^2 \leq r_{\rm j}^2$ on the $z = 0$ plane with a
radius of $r_{\rm j} = 3 \kpc$. The initial jet velocity in the nominal case, the case we will present here, is $v_{\rm j} = 8200 \kms$, a Mach number of about $10$. The direction of
the velocity at each injection point $(x,y,0)$ in the circle is
$\hat{v} = (x, y, h_{\rm j}) / \sqrt{x^2 + y^2 + h_{\rm j}^2}$,
where $h_{\rm j} = r_{\rm j} / \tan \theta_{\rm j}$. The jet is
injected during each active episode, and when the jet is turned
off reflective boundary conditions apply in the whole $z = 0$
plane. The power of the two jets together during each on-episode is
$\dot E_{2{\rm j}} = 2 \times 10^{45} \erg \s^{-1}$.
The mass deposition rate is thus
$ \dot{M}_{2{\rm j}} = {2 \dot E_{2{\rm j}}}/{v_{\rm j}^2} = 94
M_{\odot}~\yr^{-1}$.
The jet is periodically turned on for $10\myr$ and off for
$10\myr$.

The simulation begins with an isothermal box of gas at an initial
temperature of $T_{\rm ICM} (0) = 3 \times 10^7 \K$ with a density
profile of (e.g., \citealt{VernaleoReynolds2006})
\begin{equation}
\rho_{\rm ICM}(r) = \frac{\rho_0}{\left[ 1 + \left( r / a \right)
^ 2 \right] ^ {3 / 4}},
\end{equation}
with $a = 100 \kpc$ and $\rho_0 = 10^{-25} \g \cm^{-3}$. A gravity
field is added to maintain an initial hydrostatic equilibrium, and
is kept constant in time.
We include radiative cooling in the simulations, where the tabulated cooling function is taken from the solar-metalicity values of Table 6 in \cite{SutherlandDopita1993}.

% ==========================================================
\section{VORTICES}
\label{sec:vortices}
% ==========================================================

Vortices that are induced by the jets and the jet-inflated bubbles play a crucial role in the interaction of the jets with the ICM \citep{Sokeretal2013}. In particular, they lead to mixing of hot bubble gas with the ICM. This mixing heats the ICM in a process termed mixing-heating. To demonstrate vortices, we present only two examples from \cite{HillelSoker2016}; figures 4 and 5 there. More details can be found there, as well as more cases in earlier papers (e.g., \citealt{GilkisSoker2012}; \citealt{HillelSoker2014}).

\begin{figure}
\centering
\subfigure{\includegraphics[width=0.23\textwidth]{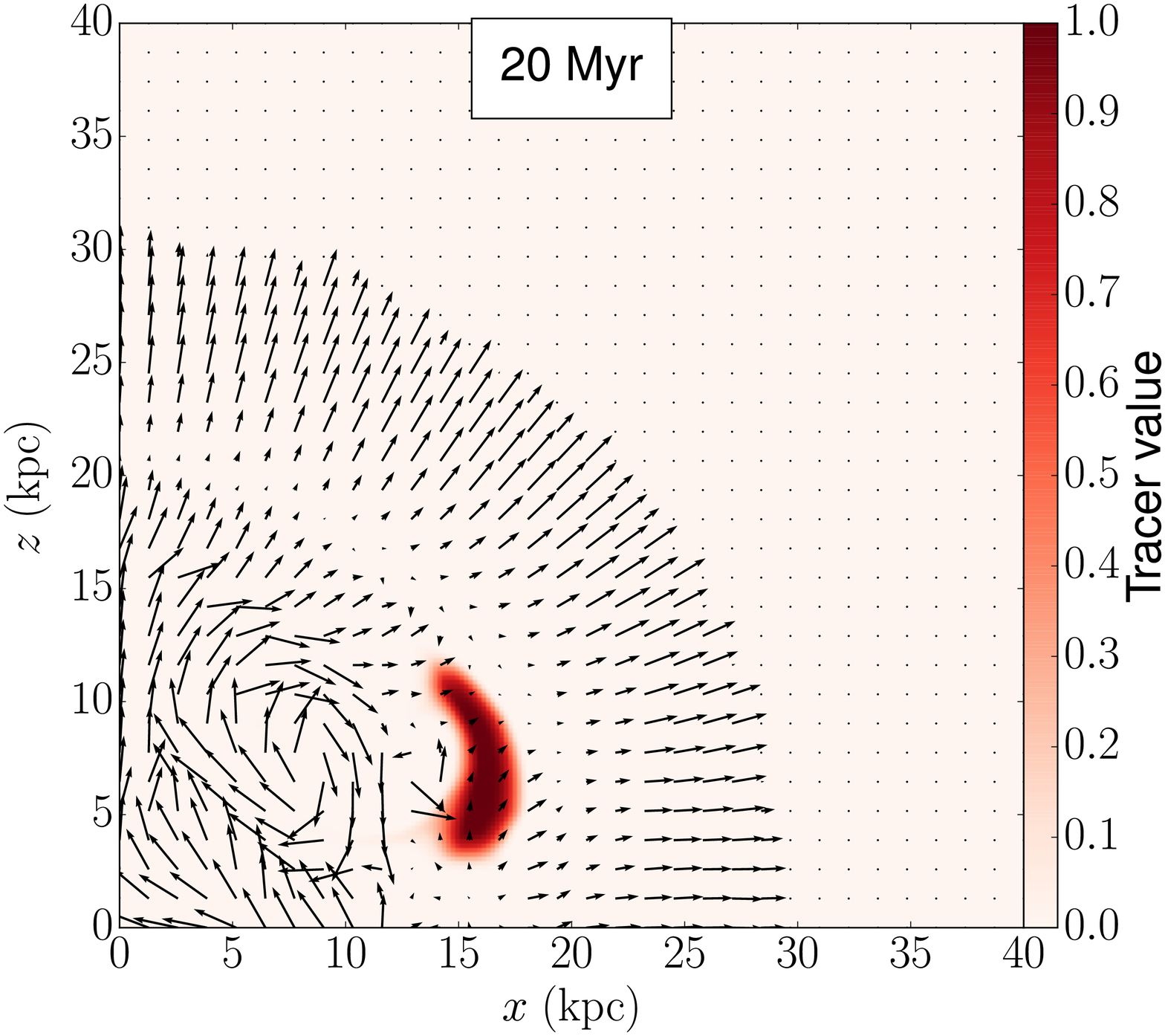}}
\subfigure{\includegraphics[width=0.23\textwidth]{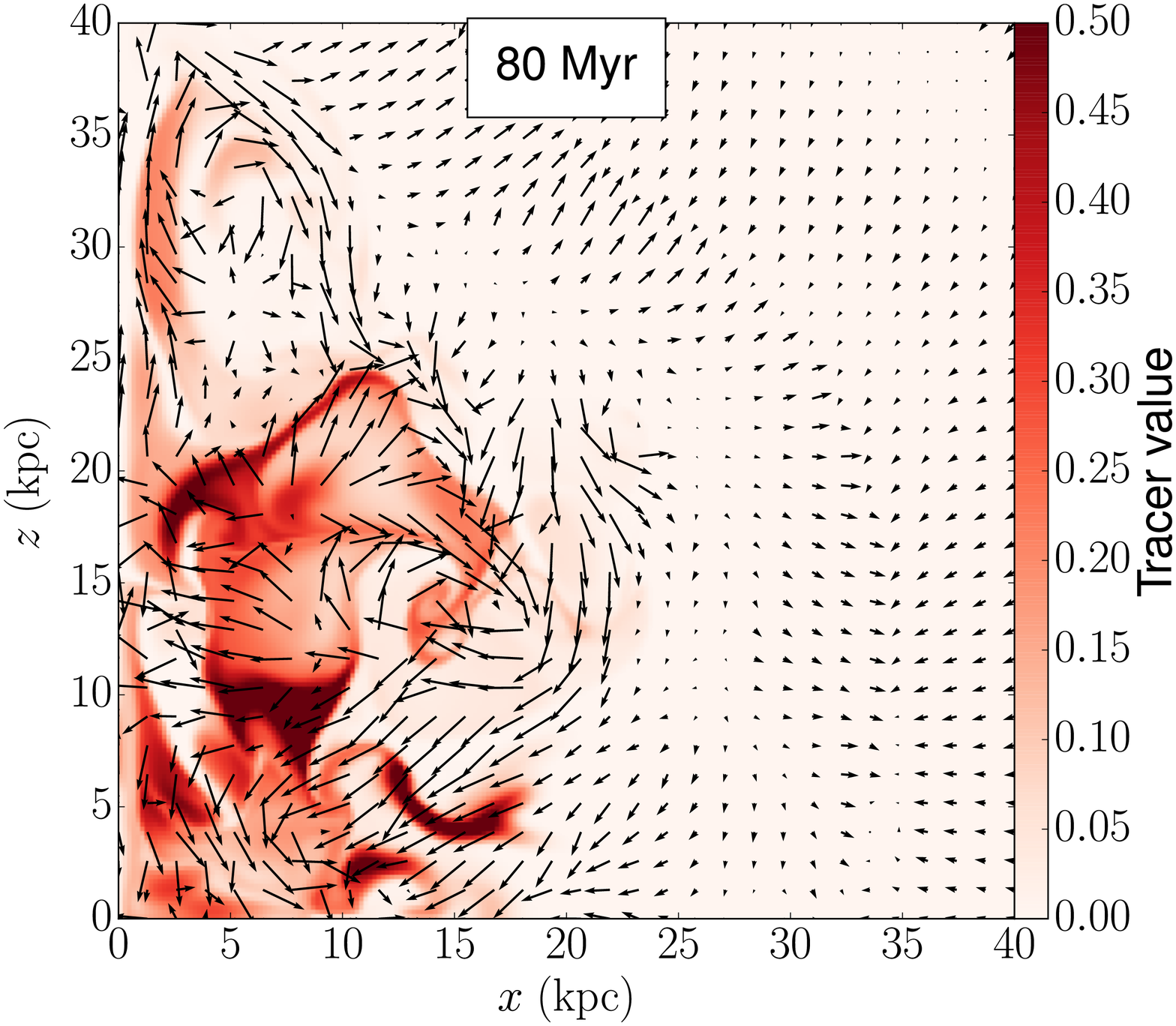}}
\caption{Flow pattern and concentration of $\tracerfour$ in the meridional plane $y=0$ at two times; the jet is injected along the $z$ axis.
Shown is the concentration of gas that at $t=0$ was contained in a torus whose axis is the $z$-axis and whose cross section is a circle centered at $(x_c,z_c)_{\rm tr}=(10,5)\kpc$ with a radius
of $r_{\rm tr}=2.5\kpc$. The largest velocity vector corresponds to $v_{\rm m}=400 \kms \simeq 0.5
{\mathcal{M}_{\rm ICM}}$, where ${\mathcal{M}_{\rm ICM}}$ is the
Mach number in the ICM. Higher velocities are marked with arrows with the same length as
that of $v_{\rm m}$. }
 \label{figure: tracers tr4}
\end{figure}

\begin{figure}
\centering
\subfigure{\includegraphics[width=0.23\textwidth]{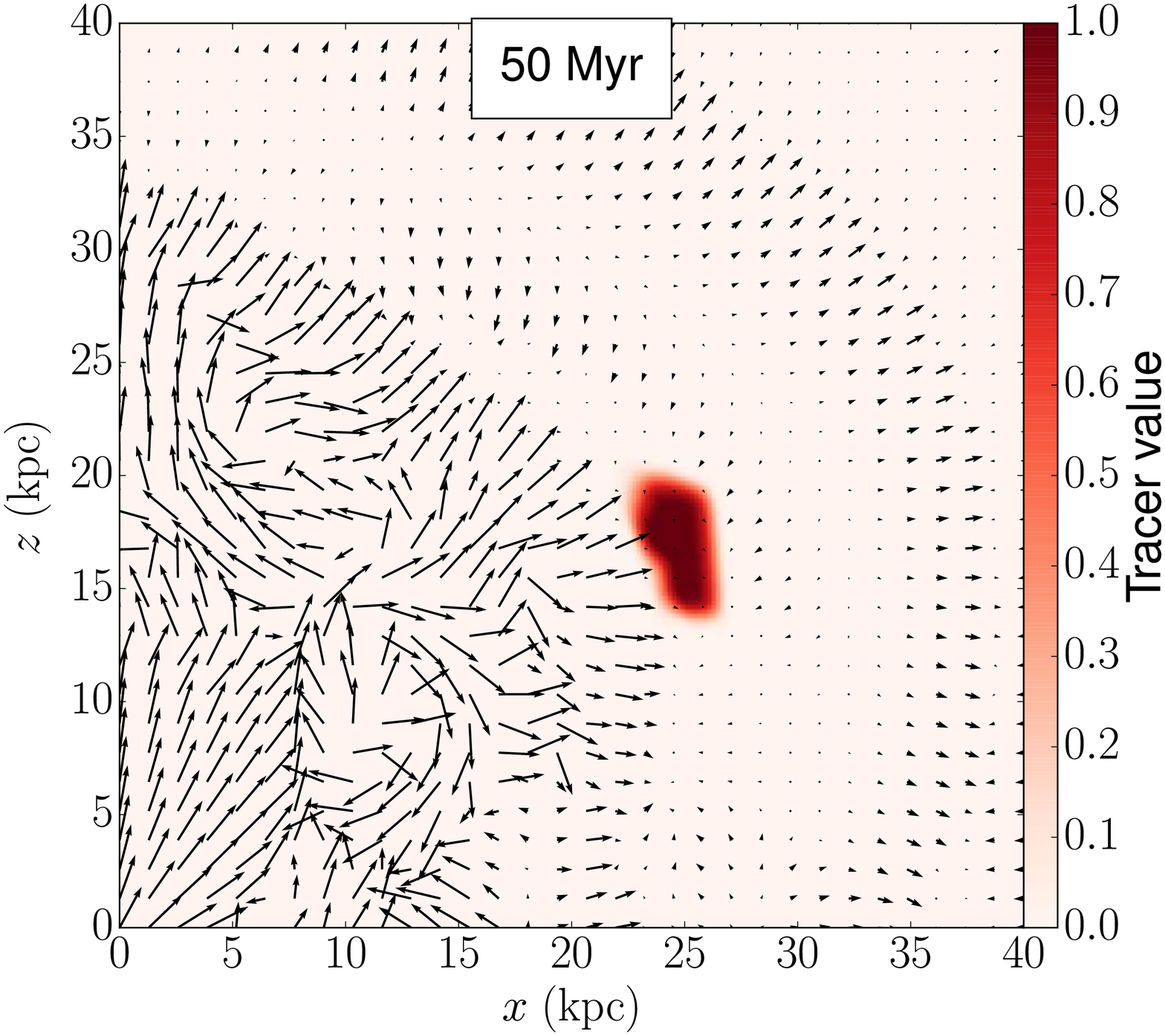}}
\subfigure{\includegraphics[width=0.23\textwidth]{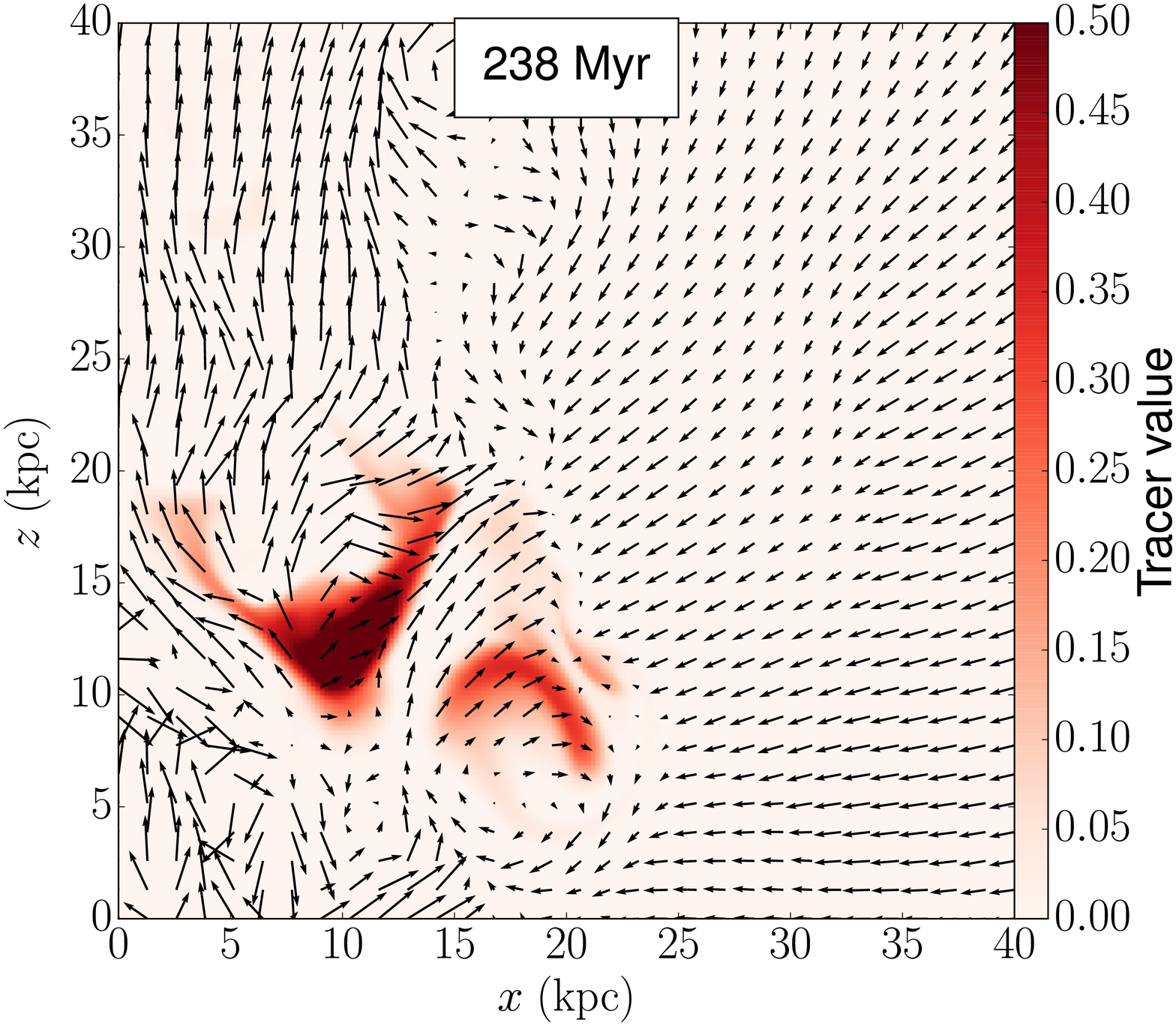}}
\caption{Like figure \ref {figure: tracers tr4} but for $\tracerthree$ that was located at $t=0$ in a torus with $(x_c,z_c)_{\rm tr}=(20,15)\kpc$ with a radius of $r_{\rm tr}=2.5\kpc$
}
 \label{figure: tracers tr3}
\end{figure}

In Figs.~\ref{figure: tracers tr4} and \ref{figure: tracers tr3} we present the flow pattern in the meridional plane (perpendicular to the symmetry plane) at two times in each figure, together with the concentration of traced gas. The scale of the flow arrows and the initial location of the tracer are described in the figure captions.
The tracers are artificial flow quantities frozen-in to the flow, and hence represent the spreading with time of gas starting in a certain volume. A tracer's initial value is set to $\xi (0) = 1$ in a certain volume and $\xi (0) = 0$ elsewhere. If the traced gas is mixed with the ICM or the jet's
material, its value drops to $0 < \xi(t) < 1$.
%FFFFFFFFFFFFFFFFFFFFFFFFFFFFFFFFFFFFFFFFFFFFFFFFFFF
%\begin{figure}[!htb]
%FFFFFFFFFFFFFFFFFFFFFFFFFFFFFFFFFFFFFFFFFFFFFFFFFFF
%FFFFFFFFFFFFFFFFFFFFFFFFFFFFFFFFFFFFFFFFFFFFFFFFFFF
%FFFFFFFFFFFFFFFFFFFFFFFFFFFFFFFFFFFFFFFFFFFFFFFFFFF

Figs.~\ref{figure: tracers tr4} and \ref{figure: tracers tr3} demonstrate the complicated flow pattern with many vortices, that is induced by the jets and the bubbles, and the vigorous  mixing that takes place.
The vortices also induce sound waves \citep{SternbergSoker2009} and turbulence \citep{HillelSoker2016}. We now turn to examine the velocity dispersion of this complicated flow.

% ==========================================================
\section{KINETIC ENERGY AND VELOCITY}
\label{sec:kinetic}
% ==========================================================

Although we did not aim to study the Perseus cluster in the simulations we present here (conducted in 2015 before the Hitomi observations), the parameters used here are not too different than those of Perseus.
The initial temperature in our simulations is $T_{\rm ICM} (0) = 3 \times 10^7 \K$,
just a little lower than the ICM temperature in Perseus \citep{Fabianetal2016}. The density in the inner region is as observed in Perseus, $n_e \approx 0.05 \cm^{-3}$ for the electron number density.
The average power of the two jets in the present simulations is $10^{45} \erg \s^{-1}$,
while that of the jets in Perseus, as inferred from the cavities \citep{Birzanetal2004, Raffertyetal2006}, is $\approx 10^{44} \erg \s^{-1}$. So we might even overestimate the velocity dispersion in Perseus. We do note that \cite{Fabianetal2016} require that the power in sound waves in their suggested heating mechanism be $6-8 \times 10^{44} \erg \s^{-1}$ at r=$10 \kpc$, not much different than the power of the jets in our simulation.

We follow the evolution of the line of sight root mean square (RMS) of the velocity of the two tracers and of the gas in the grid.
The RMS numerical velocity along the line of sight, termed here numerical velocity dispersion, is given by
\begin{equation}
\sigma_{\rm n} = \frac{\sqrt{<v^2>}}{\sqrt{3}} = \frac {1}{\sqrt{3}} \frac{2 E_{ki}}{M_i},
\end{equation}
where $i$ stands for one of the two tracers or the grid, and $E_{ki}$ and $M_i$ are the kinetic energy and mass, respectively.
We present the line-of-sight RMS velocity for gas at $T<4.5 \times 10^7 \K$, which is basically the ICM, and for gas at $T>4.5 \times 10^7 \K$ which is the shocked jet's material. The relevant quantity to compare with Hitomi is the ICM gas.

In Figs.~\ref{figure: tracers tr4} and \ref{figure: tracers tr3}
we show the evolution of the line-of-sight RMS velocity for $\tracerfour$ and $\tracerthree$ that were defined in Figs. \ref{figure: tracers tr4}, and in Fig. \ref{figure: tracers tr3}. In Fig. \ref{figure:tracergrid} we show the numerical velocity dispersion for the entire gas in the grid, according to its temperature.
In calculating the numerical velocity dispersion of the gas in the grid we consider all the gas that is located inside the numerical grid at a given time. We do not count gas that has already left the grid.
As we inject more jet's material, high velocity gas replaces lower velocity gas that has left the grid. This is the explanation for the increase in the numerical velocity dispersion in the grid that is seen in the left panel of figure \ref{figure:tracergrid}. Had we continued the simulation for a longer time, we would have reach a steady state.
%FFFFFFFFFFFFFFFFFFFFFFFFFFFFFFFFFFFFFFFFFFFFFFFFFFF
%\begin{figure}[!htb]
\begin{figure}
\centering
%\vskip -1.5 cm
\subfigure{\includegraphics[width=0.23\textwidth]{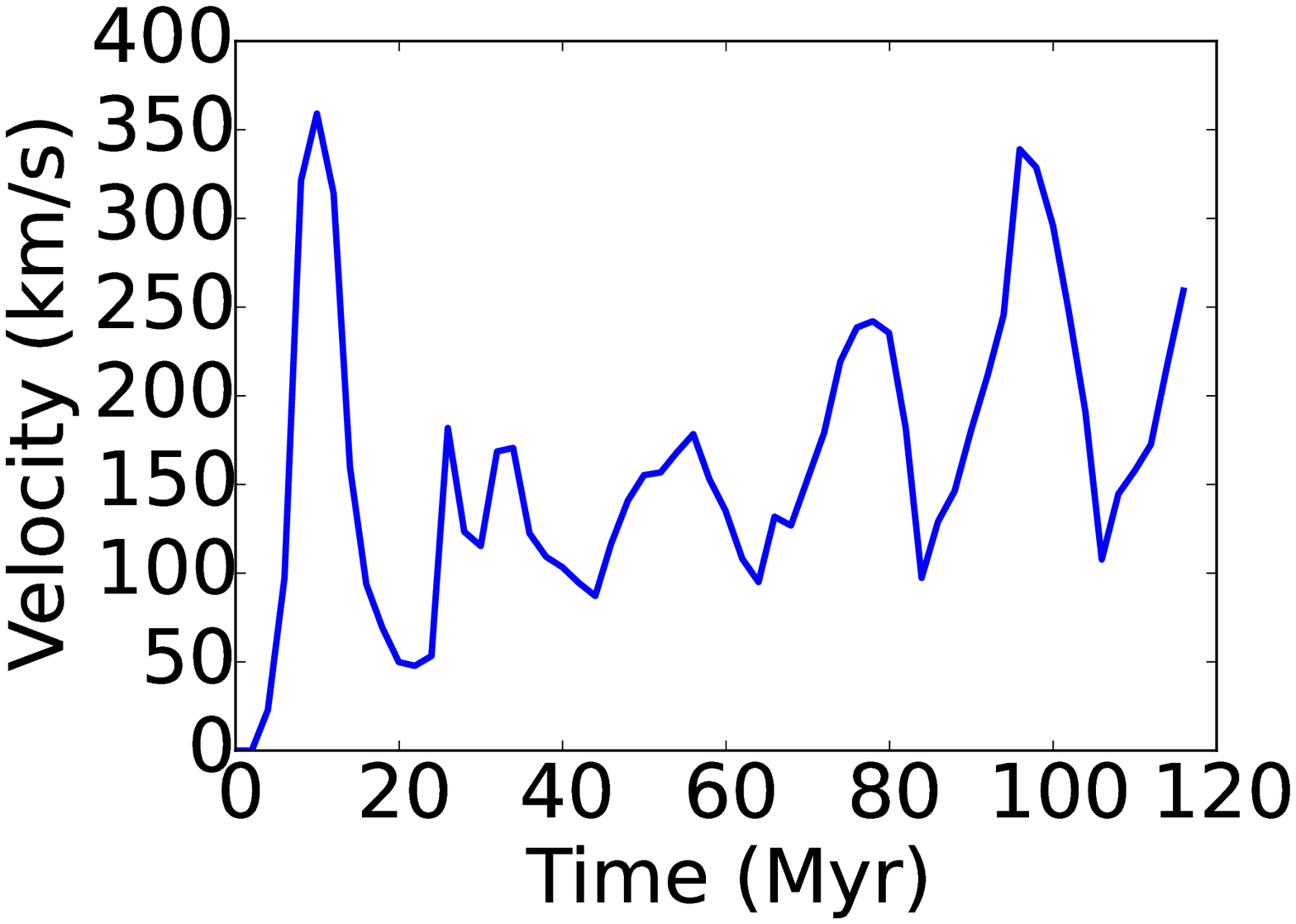}}
\subfigure{\includegraphics[width=0.23\textwidth]{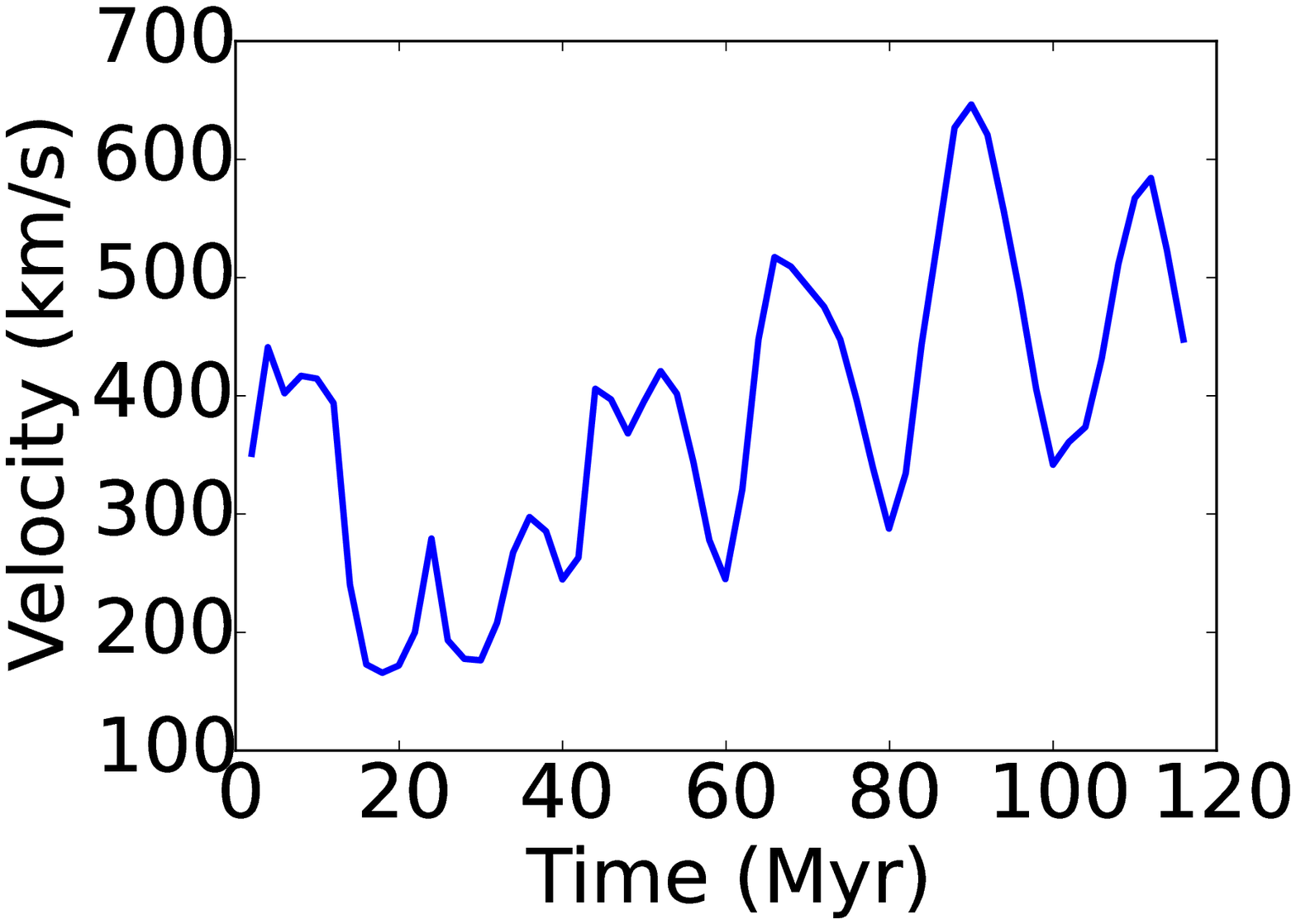}}\\
%\vskip 1.5 cm
\caption{The line-of-sight RMS velocity (numerical velocity dispersion) of $\tracerfour$, for the ICM ($T<4.5 \times 10^7 \K$; left panel), and for the hot bubble gas, i.e., the post-shock jet material ($T>4.5 \times 10^7 \K$; right panel). The left panel is relevant to the Hitomi observations. }
\label{figure:tracer4}
\end{figure}
%FFFFFFFFFFFFFFFFFFFFFFFFFFFFFFFFFFFFFFFFFFFFFFFFFFF
%FFFFFFFFFFFFFFFFFFFFFFFFFFFFFFFFFFFFFFFFFFFFFFFFFFF
%\begin{figure}[!htb]
\begin{figure}
\centering
%\vskip -1.5 cm
%\includegraphics[width =50mm]{velocity_history_below_temperature_tr3.eps}
%\vskip -0.5 cm
%\includegraphics[width=0.25\textwidth]{velocity_history_above_temperature_tr3.eps}
\subfigure{\includegraphics[width=0.23\textwidth]{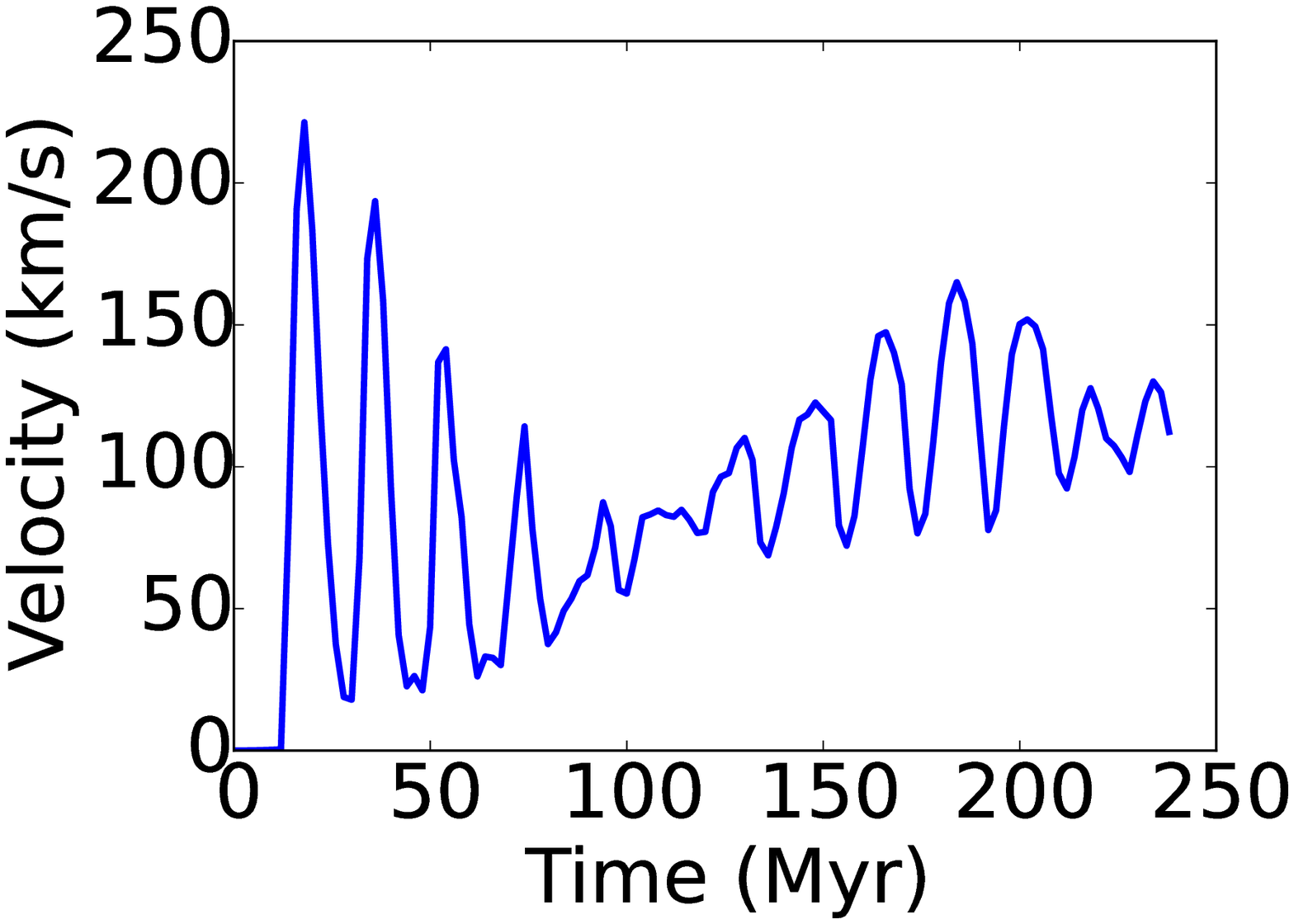}}
\subfigure{\includegraphics[width=0.23\textwidth]{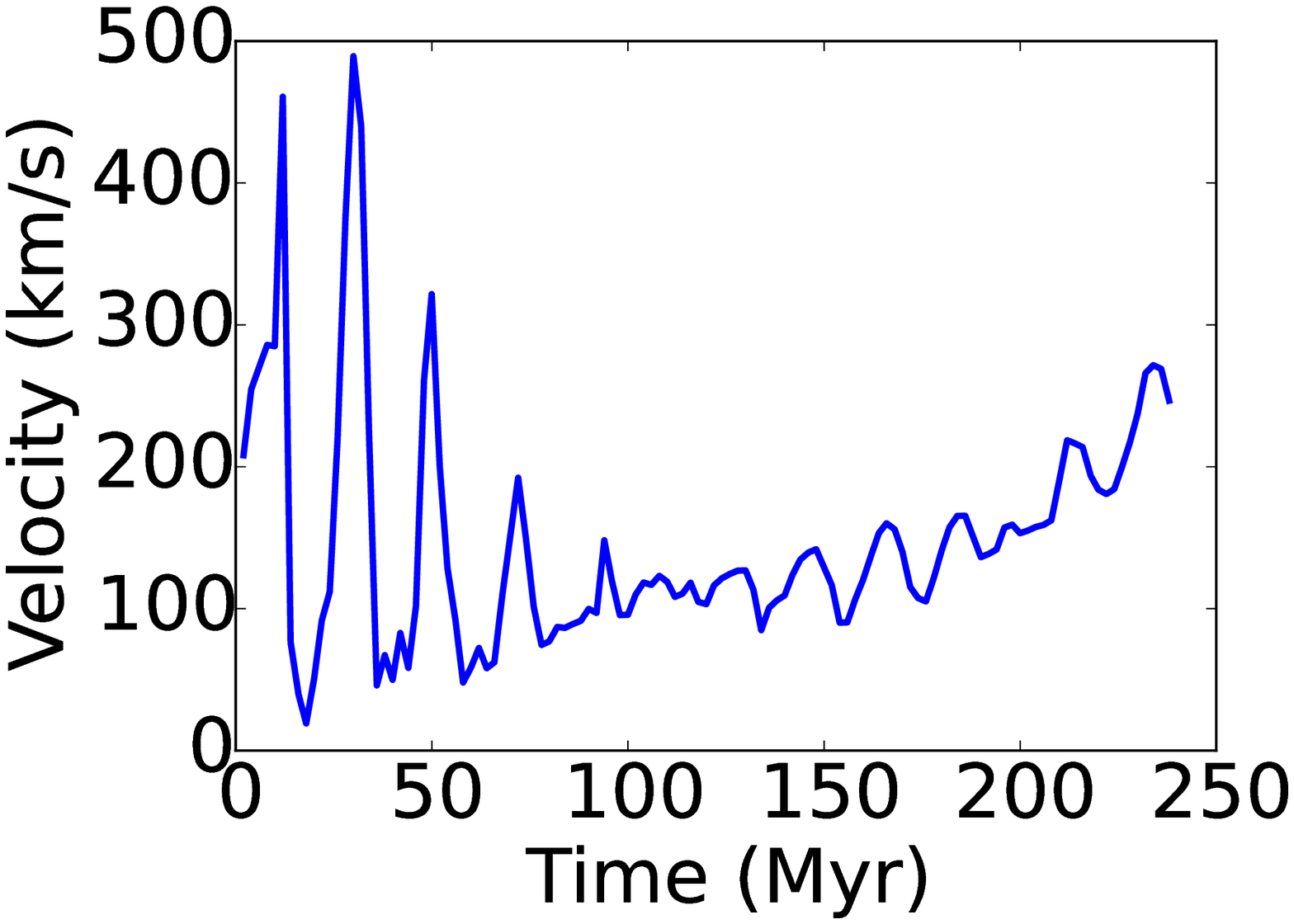}}\\
%\vskip 1.5 cm
\caption{The line-of-sight RMS velocity of $\tracerthree$, for the ICM ($T<4.5 \times 10^7 \K$; left panel), and for the hot bubble gas, i.e., the post-shock jet material ($T>4.5 \times 10^7 \K$; right panel). The left panel is relevant to the Hitomi observations. }
\label{figure:tracer3}
\end{figure}
%FFFFFFFFFFFFFFFFFFFFFFFFFFFFFFFFFFFFFFFFFFFFFFFFFFF
%FFFFFFFFFFFFFFFFFFFFFFFFFFFFFFFFFFFFFFFFFFFFFFFFFFF
%\begin{figure}[!htb]
\begin{figure}
\centering
\centering
%\vskip -1.5 cm
\subfigure{\includegraphics[width=0.23\textwidth]{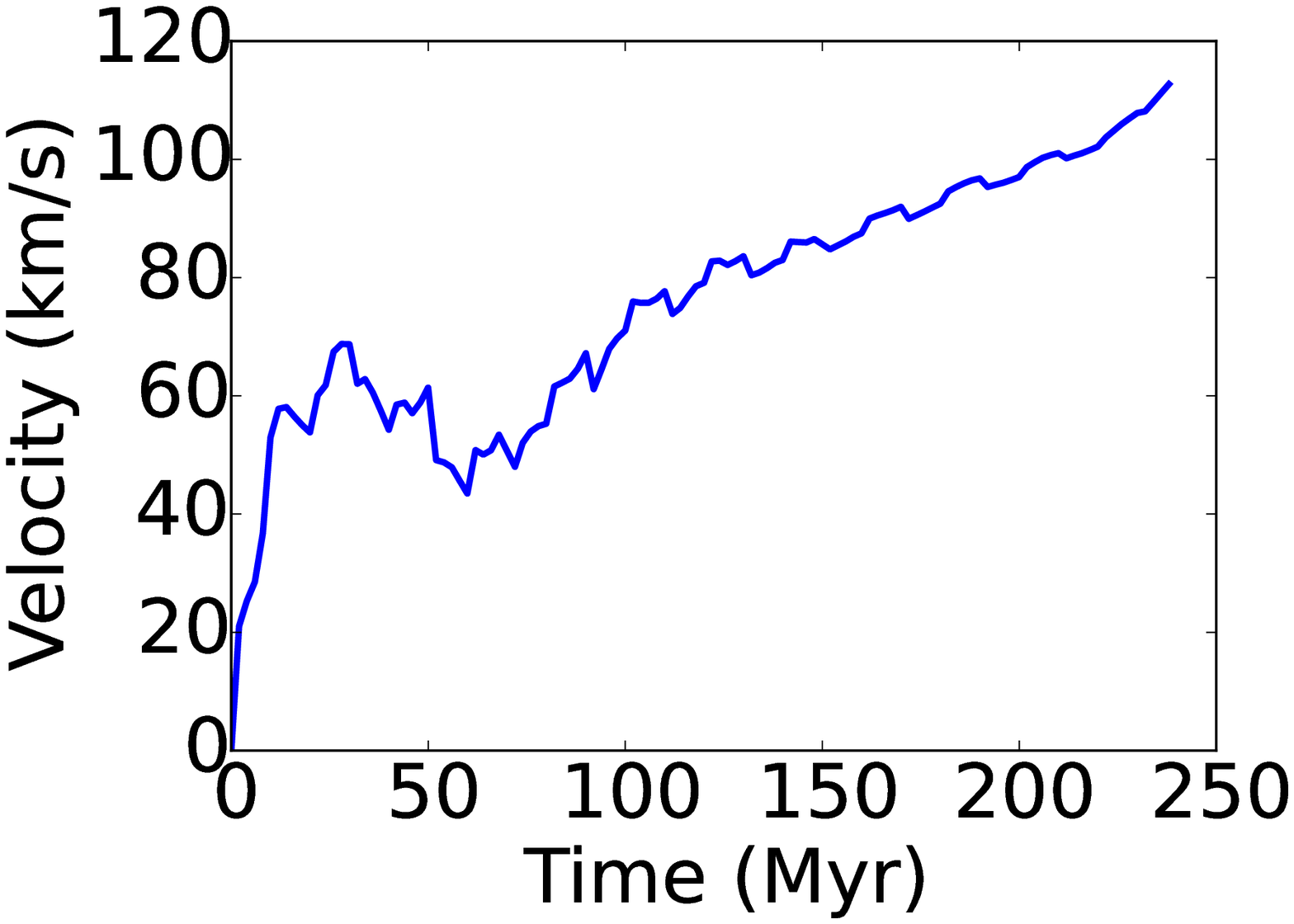}}
\subfigure{\includegraphics[width=0.23\textwidth]{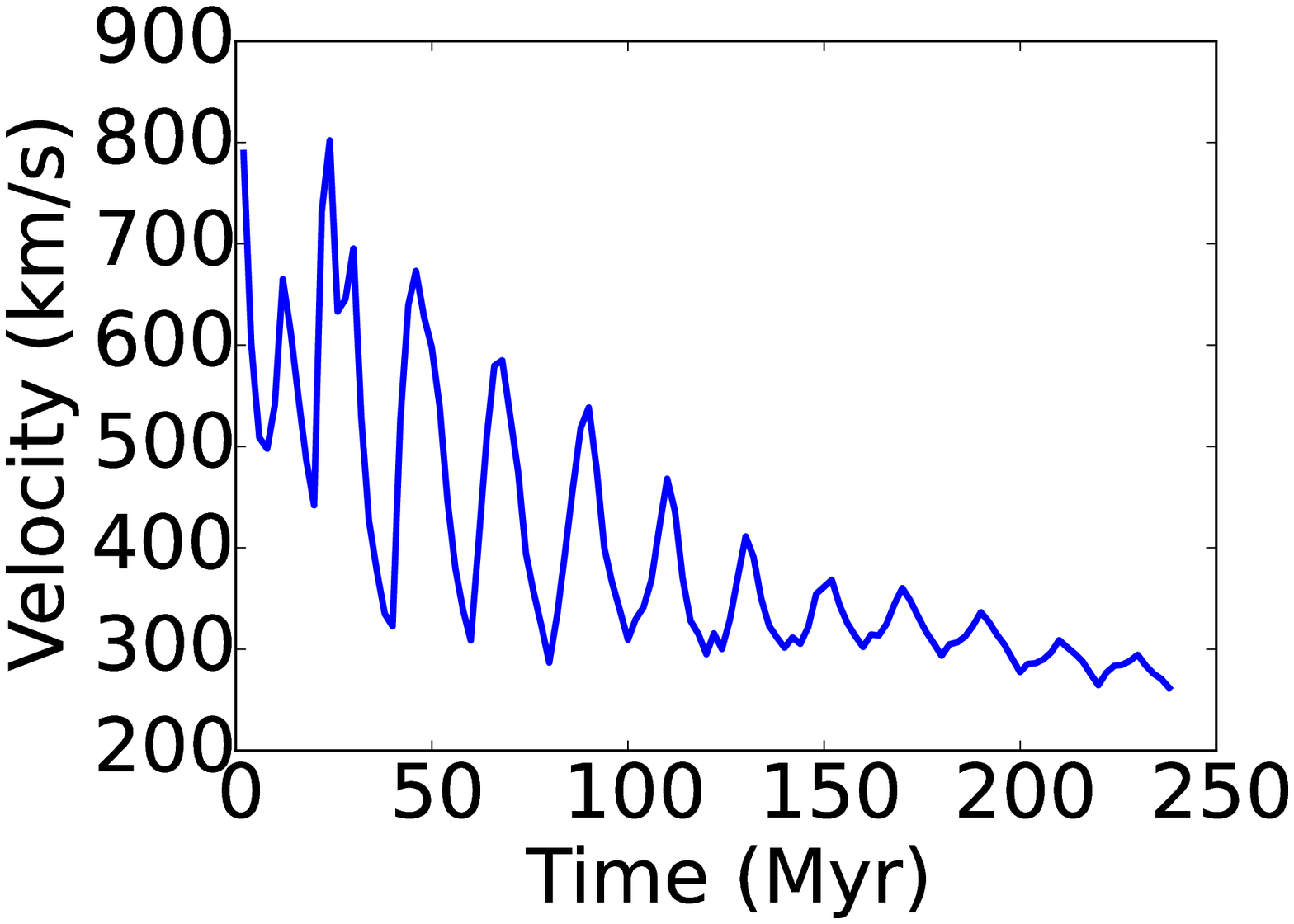}}\\
%\vskip 1.5 cm
\caption{The line-of-sight RMS velocity in the entire grid for the ICM ($T<4.5 \times 10^7 \K$; left panel), and for the hot bubble gas, i.e., the post-shock jet material ($T>4.5 \times 10^7 \K$; right panel). The left panel is relevant to the Hitomi observations. }
\label{figure:tracergrid}
\end{figure}
%FFFFFFFFFFFFFFFFFFFFFFFFFFFFFFFFFFFFFFFFFFFFFFFFFFF

For these simulation we \citep{HillelSoker2016} found that the main heating of the ICM is by mixing. Only $\approx 20\%$ of the energy is carried by kinetic energy that is distributed among shock waves, sound waves, and global flow. We find that the line-of-sight RMS velocity is in the range of $\approx 100-250 \km \s^{-1}$, compatible with the finding from Hitomi observations of a line-of-sight velocity dispersion of $164 \pm 10 \km \s^{-1}$ \citep{Hitomi2016}.
We consider this compatibility to strengthen our conclusion that heating-mixing is more significant than heating by turbulence and sound waves. Vortices that cause the heating-mixing, excite also sound waves and turbulence, and hence we expect the presence of these in the ICM.

{{{ We discuss the large fluctuations in the RMS velocity dispersion that are seen in Figs. \ref{figure:tracer3} and \ref{figure:tracer4}, and in the right panel of Fig. \ref{figure:tracergrid}. The characteristic time of these fluctuations is the period of the jet activity cycle of $20\myr$ ($10\myr$ on $10\myr$ off). Each time a tracer encounters a new jet, or the vortices it excite, it has a high value of the RMS velocity dispersion. This is even more pronounced for the hot gas in the bubble, as seen in the right panel of Fig. \ref{figure:tracergrid}. The reason for these large fluctuations is that each tracer started in a small volume, and hence represents a parcel of gas with a specific evolution. After it spreads the fluctuations decrease, as seen in Fig. \ref{figure:tracer4} for tracer B. When a large volume is included, as in the left panel of Fig. \ref{figure:tracergrid}, these fluctuations are smeared out and become very small. The observations of the Hitomi satellite include a large volume, and hence these fluctuations are expected to be small, and will weakly depend on the phase of the jet activity cycle.  }}}   

% ==========================================================
\section{SUMMARY}
\label{s-summary}
% ==========================================================

The Hitomi X-ray observations of the Perseus cluster show that dissipation of turbulent energy cannot be the main heating mechanism of the ICM. This result, that was anticipated by some studies, leaves three potential heating mechanisms of the ICM in the Perseus cluster of galaxies. These heating processes are the dissipation of sound waves \citep{Ruszkowskietal2004, Faibanetal2005, Fabianetal2016}, cosmic rays (e.g., \citealt{GuoOh2008, Fujitaetal2013}), and mixing-heating, where hot gas from jet-inflated bubbles is mixed with the ICM \citep{HillelSoker2016}. We note that heating by cosmic rays and mixing-heating are not excluded, as \cite{Pfrommer2013} argued that mixing is essential for the heating by cosmic rays model to account for observations.
Vortices that are formed in the interaction of the jets with the ICM excite both the sound waves \citep{SternbergSoker2009} and cause the mixing \citep{GilkisSoker2012, HillelSoker2014, HillelSoker2016}.

In a previous paper \citep{HillelSoker2016} we found that the main heating mechanism of the ICM is mixing-heating. The kinetic energy deposited to the ICM by the jets and the bubbles they inflate is about one quarter of the energy deposited by mixing. The kinetic energy includes also the global flow of the gas, such that the energy carried by sound waves and turbulence is less than $25 \%$ of the thermal energy deposited to the ICM by mixing.
Then there are still uncertainties about the dissipation rate of sound waves in the ICM \citep{Fabianetal2016}. If the dissipation length is large, then only a fraction of the energy that is carried by the sound waves will be dissipated in the cooling flow region in Perseus.
On the other hand, \cite{FujitaSuzuki2005} argued that the dissipation length of sound waves is too short, and the energy will be deposited only in the central region of the cooling flow.

We can summarize our view on the heating process of the ICM in Perseus, based on our numerical simulations and the new result of Hitomi, in the following way.
The inflation of bubbles by jets in the ICM is accompanied by shocks that run through the ICM and by many vortices inside and outside the jet-inflated bubbles. The vortices excite sounds waves and turbulence, and lead to mixing of hot post-shock jets' material, the hot bubble gas, with the ICM.
All four processes, shocks, dissipation of sound waves, dissipation of turbulence, and mixing, contribute to the heating of the ICM. Our finding is that mixing-heating contributes the most, more than $\approx 80 \%$ of the heating, including mixing of cosmic rays. The rest, less than $\approx 20\%$, and possibly more likely $\la 10\%$, is contributed by the other three processes that must be present, as they are by products of the inflation of bubbles.

For computational resources we acknowledge the LinkSCEEM/Cy-Tera project, which is co-funded by the European Regional Development Fund and the Republic of Cyprus through the Research Promotion Foundation.

\label{lastpage}
\end{document}